\documentclass[twocolumn,a4paper,10pt]{article}

\usepackage{arw13}

\usepackage{hyperref}

\usepackage{amssymb}
\usepackage{amsfonts,amsmath,latexsym,stmaryrd}

\usepackage{etex, xy}
\xyoption{all}

\begin{document}

 \title{Statistical Proof Pattern Recognition: Automated or Interactive?\thanks{The work was supported by EPSRC grant EP/J014222/1.}}

\institute{
	School of Computing, University of Dundee, UK
    \email{\{jonathanheras,katya\}@computing.dundee.ac.uk}
}

\author{
	J\'onathan Heras
\and
Ekaterina Komendantskaya
}



\abstract{In this paper, we compare different existing approaches employed in data mining of big proof libraries in automated and interactive theorem proving.}

\maketitle




\section{Motivation}

Over the last few decades, theorem proving has seen major developments.
\emph{Automated (first-order) theorem provers (ATPs)} (e.g. E, Vampire, SPASS) and SAT/SMT solvers
(e.g. CVC3,  Yices, Z3) are becoming increasingly fast and efficient. \emph{Interactive (higher-order) theorem provers (ITPs)} (e.g. Coq,
Isabelle/HOL, AGDA, Mizar) have been enriched with dependent types, 
(co)inductive types, type classes and provide a very rich programming environment.

The main conceptual difference between  ATPs and ITPs lies in the styles of proof development. 
For ATPs, the proof process is primarily an automatically performed \emph{proof search} in \emph{first-order} language. 
In ITPs, the proof steps are suggested by \emph{the user}, who guides the prover by providing the tactics.
ITPs work with  \emph{higher-order} logic and type theory, where many algorithms 
and procedures 
are inherently undecidable. 

Communities working on development, implementation and applications of ATPs and ITPs have
accumulated big corpora of electronic proof libraries. However, the size of the libraries, as well as 
their technical and notational sophistication often stand on the way of efficient knowledge re-use.
Very often, it is easier to start a new library from scratch rather than search the existing proof libraries
for potentially common heuristics and techniques.
Proof-pattern recognition is the area where statistical machine-learning is likely to  make an impact. 
Here, we discuss and compare two different styles of proof-pattern recognition. 

In the sequel, we will use the following convention: the term ``goal'' will stand for a an unproven proposition in the language of a given theorem prover; 
the term ``lemma'' will refer to an already proven proposition in the library.  

\section{Proof-pattern recognition in ATPs}
Given a proof goal, ATPs apply various lemmas to rewrite or simplify the goal until it is proven.
The order in which different lemmas are used plays a big role in speed and efficiency of the automated proof search. 
Hence, machine-learning techniques can be used to 
improve the premise selection procedure on the basis
of previous experience acquired from successful proofs; cf.~\cite{ku12,UrbanSPV08} .

The technical details of such machine-learning solutions would differ~\cite{KuhlweinLTUH12,TsivtsivadzeUGH11,Urban06,UrbanSPV08},
but we can summarise the common features of this approach, as follows:

\textbf{1. Feature extraction:}\\
 $\bullet$ The features are extracted from first-order formulas (given by lemmas and goals). 
 For every proposition (goal or lemma), the associated binary feature vector records, for every symbol and term of the library,
 whether it is present or absent in the proposition. As a result, the feature vectors grow to be as long as $10^6$ features long.\\
  $\bullet$ After the features are extracted, a machine-learning tool constructs a classifier for every lemma of the library, on the basis of the examples given by the feature vectors.
 For two lemmas $A$ and $B$, if $B$ was used in the proof of $A$, a feature vector $|A|$ is sent as a positive example to the classifier $<~B~>$, else $|A|$ 
 is considered to be a negative example. 
 
\textbf{2. Machine-learning tools:}\\ 
 $\bullet$ Every classifier $<B>$ has its set of positive and negative examples, hence \emph{supervised learning} is used for training.\\
 $\bullet$ The classifier algorithms \cite{KuhlweinLTUH12,TsivtsivadzeUGH11,Urban06,UrbanSPV08} range from SVMs with various kernel functions to Naive Bayes learning.	\\
 $\bullet$ Feature vectors are too big for traditional machine-learning algorithms to tackle, and the special software \emph{SNoW} is used to deal with the over-sized feature vectors.\\
 $\bullet$ The output of machine-learning algorithm provides a ``rank'' of formula lying in the interval $[0,1]$, where increasing values means increasing probability that 
$B$ is used in the proof of $A$.

\textbf{3. The mode of interaction between the prover and machine-learning tool:}\\
 $\bullet$ Given a new goal $G$, the feature vector $|G|$ is sent to the previously trained classifier $<~L~>$, for every Lemma $L$ of the given library. 
The classifier $<L>$ then outputs a rank showing how useful lemma $L$ can be in the proof of $G$.\\
 $\bullet$ Once the ranking is computed, it is used to decide, for every lemma in the library, whether it should be used in the new proof.

\textbf{4. Main improvement:} the number of goals proven automatically increases by up to 20\% - 40\%, depending on the prover and the library in question.  

Note that, if an ITP uses ATP tools to speed up the proof of first-order lemmas, the method above can be used to speed
up the automated proof search,~\cite{ku12,UrbanSPV08}.
The following figure shows this scheme of using machine-learning in ATPs and ITPs:

$$
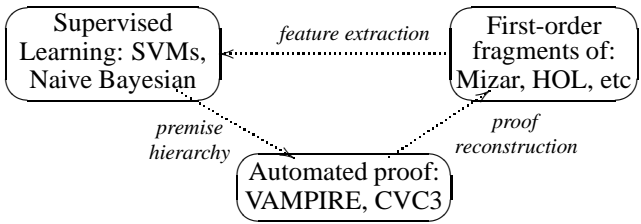
\xy0;/r.12pc/: 
(-90,0)*[o]=<80pt,35pt>\hbox{\txt{Supervised\\ Learning: SVMs,\\ Naive Bayesian}}="b"*\frm<8pt>{-},
(-27,5)*[o]=<50pt,10pt>\hbox{\txt{\footnotesize{\emph{feature extraction}}}}="a",
(-30,-35)*[o]=<80pt,25pt>\hbox{\txt{Automated proof:\\ VAMPIRE, CVC3}}="e"*\frm<8pt>{-},
(22,0)*[o]=<70pt,35pt>\hbox{\txt{First-order\\ fragments of:\\ 
Mizar, HOL, etc}}="c"*\frm<8pt>{-},
(-70,-23)*[o]=<60pt,10pt>\hbox{\txt{\emph{\footnotesize{premise}}\\ \emph{\footnotesize{hierarchy}}}}="f",
(15,-21)*[o]=<60pt,10pt>\hbox{\txt{\emph{\footnotesize{proof}}\\ \emph{\footnotesize{reconstruction}}}}="d",
"c";"b" **\dir{.} ?>*\dir{>},
"b";"e" **\dir{.} ?>*\dir{>},
"e";"c" **\dir{.} ?>*\dir{>},
\endxy
$$


\section{Proof-pattern recognition in ITPs}
Interactive style of theorem proving differs significantly from that of ATPs. In particular, a given ITP will necessarily depend on user instructions (e.g. in the form of tactics).
Because of  the inherently  interactive nature of proofs in ITPs, user interfaces play an important role in
the proof development. 
In this setting, machine-learning algorithms need to gather statistics from the user's
behaviour, and feed the results back to the user \emph{during} the proof process.
Proof-pattern recognition must become an integral part of the user interface. The first tool achieving this is ML4PG~\cite{HK12}.

Similar interfacing trend exists in the machine learning community. As statistical methods require 
users to constantly interpret and monitor results 
computed by the statistical tools, the community has developed  uniform
interfaces (\emph{Matlab, Weka}) -- environments in which the user can choose which algorithm to use for processing the data and for interpreting results.
ML4PG integrates a range of machine-learning algorithms provided by Matlab and Weka into the \emph{Proof General} -- a
 general-purpose, emacs-based interface for a range of higher-order
theorem provers.

Comparing with the ATP-based machine-learning tools, ML4PG can be characterised as follows:

 \textbf{1. Feature extraction:}\\
 $\bullet$ The features are extracted directly from higher-order propositions and proofs.\\
 $\bullet$ Feature extraction is built on the method of \emph{proof-traces}: the structure of the higher-order proposition is captured by analysing
 several proof steps the user takes when proving it, this includes the statistics of tactics, tactic arguments, tactic argument types, top symbols 
 of formulas and number of generated subgoals, see \cite{HK12}. \\
$\bullet$ The feature vectors are fixed at the size of 30. This size is manageable for literally any 
 existing statistical machine-learning algorithm.\\
  $\bullet$ Longer proofs are analysed by means of the \emph{proof-patch} method: when features of one big proof are collected by taking a collection of
 features of smaller proof fragments.

\textbf{2. Machine-learning tools:}\\ 
 $\bullet$ As higher-order proofs in general can take a variety of shapes, sizes and proof-styles, ML4PG does not use
any \emph{a priori} given training labels. Instead, it uses unsupervised learning (\emph{clustering}), and in particular, Gaussian, k-means, and 
farthest-first algorithms. \\
 $\bullet$ The output of clustering algorithm provides proof families based on some user defined parameters -- e.g. cluster size, and 
proximity of lemmas within the cluster.

\textbf{3. The mode of interaction between the prover and machine-learning tool:}\\
 $\bullet$ ML4PG works on the background of Proof General, and extracts the features interactively in the process of Coq compilation.\\
 $\bullet$ On user's request, it sends the gathered statistics to a chosen machine-learning
interface and triggers execution of a clustering algorithm of the user's choice, using adjustable user-defined clustering parameters.\\
 $\bullet$ ML4PG does some gentle post-processing of the results given by the machine-learning tool, and displays families of related proofs to the user.

\textbf{4. Main improvement:} ML4PG makes use of the rich interfaces in ITPs and machine learning. 
It assists the user, rather than the prover: the user may treat the suggested similar lemmas as proof hints.
The interaction with ML4PG is fast and easy, so the user may receive these hints interactively, and in real time. The process is summarised below:   

$$
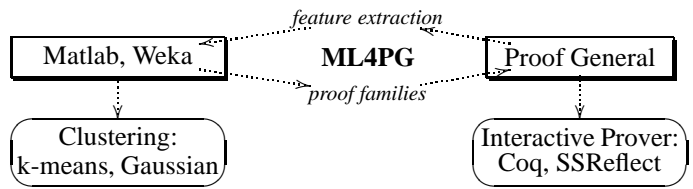
\xy0;/r.12pc/: 
(-80,0)*[o]=<80pt,15pt>\hbox{\txt{Matlab, Weka}}="b"*\frm<8pt>{-,},
(-80,-25)*[o]=<80pt,25pt>\hbox{\txt{Clustering:\\ k-means, Gaussian}}="e"*\frm<8pt>{-},
(40,0)*[o]=<70pt,15pt>\hbox{\txt{Proof General}}="c"*\frm<8pt>{-,},
(40,-25)*[o]=<80pt,25pt>\hbox{\txt{Interactive Prover:\\  Coq,  SSReflect}}="f"*\frm<8pt>{-},
(-15,10)*[o]=<60pt,10pt>\hbox{\txt{\emph{\footnotesize{feature extraction}}}}="a",
(-15,-10)*[o]=<60pt,10pt>\hbox{\txt{\emph{\footnotesize{proof families}}}}="d",
(-15,0)*[o]=<60pt,10pt>\hbox{\txt{\textbf{ML4PG}}}="g",
"c";"a" **\dir{.} ?>*\dir{>},
"a";"b" **\dir{.} ?>*\dir{>},
"b";"d" **\dir{.} ?>*\dir{>},
"d";"c" **\dir{.} ?>*\dir{>},
"b";"e" **\dir{.} ?>*\dir{>},
"c";"f" **\dir{.} ?>*\dir{>},
\endxy
$$

\section{Conclusions}
The automated and interactive styles of proof-pattern recognition described here have both been successfully 
applied in big proof libraries in Mizar, HOL, Isabelle, Coq, and SSReflect. The methods complement each other: one aims to speed up the first-order proofs, and the other one
provides guidance where proofs cannot be fully automated.

\bibliographystyle{plain}
\bibliography{mlipgii,katya2}

\end{document}